\documentclass[aps,prd,twocolumn,showpacs,superscriptaddress]{revtex4}

\usepackage{graphicx}
\usepackage{epsfig}
\usepackage{psfrag}
\usepackage{bbm}

\begin{document}

\title{Local readout enhancement for detuned signal-recycling interferometers}

\author{Henning Rehbein}
\author{Helge M\"uller-Ebhardt}
\affiliation{Max-Planck-Institut f\"ur Gravitationsphysik
             (Albert-Einstein-Institut), \\ Institut f\"ur Gravitationsphysik,
              Leibniz Universit\"at Hannover, Callinstr. 38, 30167 Hannover,
              Germany}
\author{Kentaro Somiya}
\affiliation{Max-Planck-Institut f\"ur Gravitationsphysik (Albert-Einstein-Institut),
             Am M\"uhlenberg 1, 14476 Potsdam, Germany}
\author{Chao Li}
\affiliation{California Institute of Technology, M/C 130-33,
             Pasadena, CA 91125, USA}
\author{Roman Schnabel}
\author{Karsten Danzmann}
\affiliation{Max-Planck-Institut f\"ur Gravitationsphysik
             (Albert-Einstein-Institut), \\ Institut f\"ur Gravitationsphysik,
              Leibniz Universit\"at Hannover, Callinstr. 38, 30167 Hannover,
              Germany}
\author{Yanbei Chen}
\affiliation{Max-Planck-Institut f\"ur Gravitationsphysik (Albert-Einstein-Institut),
             Am M\"uhlenberg 1, 14476 Potsdam, Germany}

\date{\today}

\begin{abstract}
High power detuned signal-recycling interferometers currently
planned for second-generation interferometric gravitational-wave
detectors (for example Advanced LIGO) are characterized by two
resonances in the detection band, an optical resonance and an
optomechanical resonance which is upshifted from the suspension
pendulum frequency due to the so-called {\it optical-spring}
effect. The detector's sensitivity is enhanced around these two
resonances. However, at frequencies below the optomechanical
resonance frequency, the sensitivity of such interferometers is
significantly lower than non-optical-spring configurations with
comparable circulating power; such a drawback can also compromise
high-frequency sensitivity, when an optimization is performed on
the overall sensitivity of the interferometer to a class of
sources. In this paper, we clarify the reason of such a low
sensitivity, and propose a way to fix this problem. Motivated by
the {\it optical-bar} scheme of Braginsky, Gorodetsky and Khalili,
we propose to add a local readout scheme which measures the motion
of the arm-cavity front mirror, which at low frequencies moves
together with the arm-cavity end mirror, under the influence of
gravitational waves. This scheme improves the low-frequency
quantum-noise-limited sensitivity of optical-spring
interferometers significantly and can be considered as a
incorporation of the optical-bar scheme into currently planned
second-generation interferometers. On the other hand it can be
regarded as an extension of the optical bar scheme. Taking
compact-binary inspiral signals as an example, we illustrate how
this scheme can be used to improve the sensitivity of the planned
Advanced LIGO interferometer, in various scenarios, using a
realistic classical-noise budget. We also discuss how this scheme
can be implemented in Advanced LIGO with relative ease.
\end{abstract}


\maketitle

\section{Introduction}

First-generation laser interferometric gravitational-wave (GW)
detectors (LIGO~\cite{SHOEMAKER2004}, VIRGO~\cite{FIORE2002},
GEO~\cite{WIL2002} and TAMA~\cite{ANDO2001}) are reaching design
sensitivities. These interferometers are usually Michelson
interferometers with Fabry-Perot cavities in the arms, with
power-recycling (PR) at the laser input port (with the exception
of GEO, which uses {\it dual-recycling}~\cite{HFGSD2002}), and
operating close to the dark-port condition.

In order to have a flexible sensitivity to specific astrophysical
sources, and for other technical reasons such as lowering power at
the beam splitter (BS), second-generation interferometers, such as
Advanced LIGO~\cite{advLIGO}, plan to use the so-called
signal-recycling (SR) configuration, in which an additional mirror
is placed at the dark port of a Fabry-Perot Michelson
interferometer, modifying the optical  resonant structure of the
interferometer. The adjustment of the location and reflectivity of
the signal-recycling mirror varies the optical resonance frequency
and bandwidth, respectively. Near the optical resonance,
sensitivity to GWs is improved. When the signal-recycling cavity,
the cavity formed by the input test-mass mirrors and the
signal-recycling mirror is neither resonant nor anti-resonant with
respect to the carrier frequency, the optical configuration is
called {\it detuned} signal-recycling. In these detuned
configurations, the optical resonance of the interferometer is
away from the carrier frequency, creating a peak sensitivity to
GWs away from DC.
\begin{figure}[h]
\includegraphics[width=6.5cm]{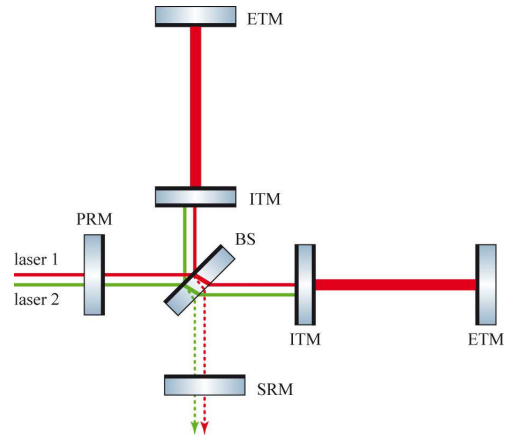}
\caption{Schematic plot of a power- and signal-recycled Michelson
interferometer with arm cavities and double-readout. The added
local readout sensing the ITM is realized by a secondary laser
which does not resonate in the arm cavities.} \label{mich}
\end{figure}

As demonstrated theoretically by Buonanno and
Chen~\cite{BUCH2001,BUCH2002,BUCH2003} and experimentally by
Somiya et al.~\cite{Som2005} and Miyakawa et
al.~\cite{Miyakawa2006}, detuned signal-recycling also makes the
power inside the interferometer depend on the motion of the
mirrors, creating an {\it optical spring}, and can shift the
eigenfrequency of the test masses from the pendulum frequency
($\sim 1\,$Hz) up to the detection band. The optical spring helps
to improve the interferometer's response to GWs around the
optomechanical resonant frequency, even allowing the
interferometer to surpass the free-mass Standard Quantum Limit
(SQL). However, the quantum-noise-limited sensitivity of
optical-spring interferometers at frequencies below the
optomechanical resonant frequency is dramatically lower than the
one of non-optical-spring interferometers. Such a limitation in
sensitivity is caused by the optical spring, which rigidly
connects the front and the end mirror of the arm cavities at
frequencies below the optomechanical resonance. The general
principle underlying this effect has already been explained in the
works of Braginsky, Gorodetsky and Khalili, namely in their
proposal of the {\it optical bar} detection scheme~\cite{BGK1997}.
In order to understand this more conveniently, we need to use the
local inertial frame of the BS, in which the effect of GWs can be
described completely as a tidal force field, which induces forces
only on the end test-mass mirrors (ETMs), but not on the input
test-mass mirrors (ITMs). We make the approximation that the ITMs
and the BS are co-located. In this frame, the propagation of the
light is unaffected by GWs. Remember that the optical spring
connects the ITM and the ETM.  At frequencies substantially below
the optomechanical resonance, the optical spring behaves like a
rigid optical bar, connecting the ITM and the ETM of each arm
rigidly. It is then easy to understand that the carrier light,
which senses the change in arm-cavity length, or the difference in
ITM and ETM motion, cannot be used to measure GW efficiently at
these frequencies.  On the other hand, since the ITM and the ETM
are rigidly connected, they both move, in the local inertial frame
of the BS, by $1/2$ the amount the ETM would have moved if there
were no optical spring present (assuming ITM and ETM to have equal
masses). To illustrate this situation, assume that a low-frequency
GW with amplitude $h$ is incident from right above our detector
(with arm-length $L$), then in the local inertial frame of the BS,
the motion of the ETM of a non-optical-spring interferometer would
be $L h$, the motion of ITM and ETM of an optical-spring
interferometer below resonance will be both $\sim Lh/2$. For this
reason, if one also measures the local motion of the ITM using an
additional {\it local readout} scheme, one can recover
low-frequency sensitivity dramatically. Note that as viewed by the
local meter, the ITM has an effective mass that is equal to the
total mass of the ITM and the ETM. If one applies a local readout
scheme to the ETM, the same sensitivity recovery is possible,
since the ETM also moves with respect to a free co-located mirror
by $-Lh/2$. Braginsky, Gorodetsky and Khalili proposed an
optical-bar detection scheme, in which only the local motion of
the ITM is measured~\cite{BGK1997}. In this sense, what we are
proposing can be considered as directly incorporating the
optical-bar scheme into currently planned second-generation
interferometers.

Local readout schemes have also been proposed  for interferometers
without optical spring, with a different motivation. In those
interferometers, the motion of mirrors with respect to their local
inertial frames are caused by radiation-pressure noise (if we only
consider signal and quantum noise sources); results of local
readout schemes can thus be used to cancel radiation-pressure
noise and improve low-frequency sensitivity
\cite{KMS2001,CoHePi2003}. Furthermore, such schemes are able to
cancel parts of the classical noise. Our treatment here can also
be viewed as a generalization of these schemes because by setting
detuning in our treatment to zero will recover their results.

From an astrophysical point of view, the addition of the local
readout scheme, which broadens the detection band, will allow the
interferometer to search for multiple sources simultaneously, as
well as to examine a wider frequency range of the same source.  As
an example, we will explore how the increase in detection
bandwidth can allow us to detect more efficiently the population
of compact binary objects with a broad range of masses (and hence
signal frequency band).

In order to construct the local meter, we consider a scheme where
a second carrier is injected into the bright port, which does not
enter the arm cavities, but instead senses the location of the
ITMs, as shown in Fig.\,\ref{mich}. An alternative strategy would
be attaching auxiliary interferometers at the ETMs. These two
strategies are quite equivalent in the ideal situation, but differ
from each other in terms of difficulty in implementation, in terms
of quantum noise and in terms of technical noise sources such as
laser noise as we will discuss in some more details.

This paper is organized as follows. In Sec.\,\ref{sec:eqm}, we
study the dynamics, sensing, and control of our double-readout
interferometer.  In Sec.\,\ref{sec:eqm:em}, we write down and
solve the joint Heisenberg equations of motion of test masses,
beam splitter, and optical fields; in Sec.\,\ref{sec:eqm:sn}, we
evaluate the optimal combined GW sensitivity of the two readout
channels; in Sec.~\ref{sec:eqm:ctrl}, we prove that the use of
control schemes do not affect this sensitivity. In
Sec.\,\ref{sec:implementation} we show the benefit, which the
local readout scheme will provide for the detection of
intermediate-mass black-holes, using a realistic Advanced LIGO
noise budget. In Sec.\,\ref{sec:implementation} we consider
practical issues for a possible implementation in Advanced LIGO.
In Sec.\,\ref{sec:conclusion} we summarize our main conclusions.

\section{Dynamics, sensing and control}
\label{sec:eqm}

\subsection{Equations of motion}
\label{sec:eqm:em}
Let us consider a configuration where the ITMs' motion of a
signal- and power-recycled Michelson interferometer with arm
cavities is locally sensed by a small interferometer which has the
ITMs as its end mirrors (cf. Fig.\,\ref{mich}). This is realized
by injecting a second carrier into the bright port, which does not
resonate in the arms (preferably anti-resonant). Because the
frequency (and the polarization) of the second carrier is (are)
different from that of the first, we effectively obtain two
interferometers in one scheme where parameters such as detuning
and mirror reflectivities for each interferometer can be chosen
independently; input vacuum fluctuations associated with the two
lasers are also independent.

Throughout this paper, we will assume the GW with amplitude $h$ as
incident from right above the interferometer, with a polarization
that maximizes the response of our $L$-shaped Michelson
interferometers.  In the following we will list the Heisenberg
equation of motions in frequency domain
\cite{KLMTV2001,BUCH2001,BUCH2002,BUCH2003,CCM2005} for the
differential mode of motion (i.e., opposite in the two arms) of
the input mirrors $\hat x_{\rm ITM}$ and the end mirrors $\hat
x_{\rm ETM}$, respectively, as well as for the BS motion normal to
its reflective surface $\hat x_{\rm BS}$ and for the two
measurement outputs $\hat y^{(i)}$
\begin{widetext}
\begin{eqnarray}
\hat x_{\rm ITM} &=& - R_{xx} (\Omega) \ \left[\hat F^{(1)}
(\Omega) + R_{FF}^{(1)} (\Omega) \ (\hat x_{\rm ETM} - \hat x_{\rm
ITM}) - \hat F^{(2)} (\Omega) - R_{FF}^{(2)} (\Omega) \ (\hat
x_{\rm ITM} + \sqrt{2} \ \hat x_{\rm BS})\right] + \hat \xi_{\rm
ITM}\,,
\label{eomitm}\\
\hat x_{\rm ETM} &=& R_{xx} (\Omega) \ \left[\hat F^{(1)} (\Omega)
+ R_{FF}^{(1)} (\Omega) \ (\hat x_{\rm ETM} - \hat x_{\rm
ITM})\right] + L \ h + \hat \xi_{\rm ETM}\,,
\label{eometm}\\
\hat x_{\rm BS} &=& R_{xx}^{\rm BS} (\Omega) \ \left[\hat F^{(2)}
(\Omega) + R_{FF}^{(2)} (\Omega) \ (\hat x_{\rm ITM} + \sqrt{2} \
\hat x_{\rm BS}) + \hat F_{\rm BP}^{(1)} (\Omega)+\hat F_{\rm
BP}^{(2)} (\Omega)\right] + \hat \xi_{\rm BS}\,,
\label{eombs}\\
\hat y^{(1)} &=& \hat Y_1^{(1)} (\Omega) \ \sin \zeta^{(1)} + \hat
Y_2^{(1)} (\Omega) \ \cos \zeta^{(1)} + \left[R_{Y_1F}^{(1)}
(\Omega) \ \sin \zeta^{(1)} + R_{Y_2F}^{(1)} (\Omega) \ \cos
\zeta^{(1)}\right] (\hat x_{\rm ETM} - \hat x_{\rm ITM})
\,, \label{eom1} \\
\hat y^{(2)} &=& \hat Y_1^{(2)} (\Omega) \ \sin \zeta^{(2)} + \hat
Y_2^{(2)} (\Omega) \ \cos \zeta^{(2)} + \left[R_{Y_1F}^{(2)}
(\Omega) \ \sin \zeta^{(2)} + R_{Y_2F}^{(2)} (\Omega) \ \cos
\zeta^{(2)}\right] (\hat x_{\rm ITM} + \sqrt{2} \ \hat x_{\rm
BS})\,. \label{eom2}
\end{eqnarray}
\end{widetext}
Note that $\hat x_{\rm ITM}$ and $\hat x_{\rm ETM}$ account for
the differential motion between two mirrors while $\hat x_{\rm
BS}$ describes the motion of a single mirror with an angle of 45
degree. This explains the factor of $\sqrt{2}$ in front of the BS
motion. The out-going fields at the dark port belonging to the two
different carriers are each sensed by homodyne detection such that
the measurement outputs are a certain combination of amplitude and
phase quadratures (described by the phases $\zeta^{(1),(2)}$).
Note that we have labeled all quantities with superscripts $(1)$
and $(2)$ for the large-scale interferometer and the local meter
(the small interferometer, formed by the BS and the ITMs),
respectively. The operators $\hat F^{(i)}$ and $\hat F^{(i)}_{\rm
BP}$ describe the radiation pressure forces which would act on
fixed mirrors caused by the incoming vacuum fields at the dark
port and the laser light fluctuations from the bright port,
respectively. The operators $\hat Y^{(i)}_j$ account for the shot
noise in case of fixed mirrors. Each optical component is subject
to classical noise generated by the corresponding operator $\hat
\xi$ and has its own mechanical susceptibility $R_{xx}$. The
susceptibilities $R^{(i)}_{FF}$ describe the optical springs
\cite{BUCH2001} and $R^{(i)}_{Y_{i}F}$ the transformation of the
mirror motion into the two outputs. In the following we will
present all these quantities more detailed while all appearing
parameters are summarized in Tab.\,\ref{definitions}.

The free radiation pressure force and the free shot noise in each
of the two interferometers are given by \cite{BUCH2003}
\begin{eqnarray}
\hat F^{(i)} &=& \sqrt{\frac{\epsilon^{(i)} \theta^{(i)}m \hbar
}{2}} \frac{({\rm i} \Omega - \epsilon^{(i)}) \ \hat a_1^{(i)} +
\lambda^{(i)} \ \hat a_2^{(i)}}{(\Omega - \lambda^{(i)} + {\rm i}
\epsilon^{(i)}) (\Omega + \lambda^{(i)}
+ {\rm i} \epsilon^{(i)})}\,, \nonumber \\
\hat Y_1^{(i)} &=& \frac{((\lambda^{(i)})^2 - (\epsilon^{(i)})^2 -
\Omega^2) \ \hat a_1^{(i)} + 2 \lambda^{(i)} \epsilon^{(i)} \ \hat
a_2^{(i)}}{(\Omega - \lambda^{(i)} + {\rm i} \epsilon^{(i)})
(\Omega + \lambda^{(i)} + {\rm i} \epsilon^{(i)})}\,,\nonumber \\
\hat Y_2^{(i)} &=& \frac{-2 \lambda^{(i)} \epsilon^{(i)} \ \hat
a_1^{(i)} + ((\lambda^{(i)})^2 - (\epsilon^{(i)})^2 - \Omega^2) \
\hat a_2^{(i)}}{(\Omega - \lambda^{(i)} + {\rm i} \epsilon^{(i)})
(\Omega + \lambda^{(i)} + {\rm i} \epsilon^{(i)})}\,, \nonumber
\end{eqnarray}
where $\theta^{(i)} = \frac{8 P^{(i)} \omega_0^{(i)}}{m L^{(i)}
c}$ has units of frequency cube. Note that $P^{(i)}$ refers to the
circulating power in each arm, respectively. Here $\hat a_1^{(i)}$
and $\hat a_2^{(i)}$ are the amplitude and phase quadrature
operators of the incoming vacuum fields at the dark port
\cite{KLMTV2001}, associated with the first and second carrier
field, respectively. The susceptibilities are given by
\cite{BUCH2003}
\begin{eqnarray}
R_{xx}^{\rm BS} &=& -\frac{\sqrt{2}}{m_{\rm BS} \Omega^2}\,,\nonumber\\
R_{xx} &=& -\frac{2}{m \Omega^2}\,,\nonumber\\
R_{FF}^{(i)} &=& \frac{\theta^{(i)} m}{4}
\frac{\lambda^{(i)}}{(\Omega - \lambda^{(i)} + {\rm
i} \epsilon^{(i)}) (\Omega + \lambda^{(i)} + {\rm i} \epsilon^{(i)})}\,,\nonumber \\
R_{Y_1F}^{(i)} &=& \sqrt{\frac{\epsilon^{(i)} \theta^{(i)} m}{2
\hbar}} \frac{\lambda^{(i)}}{(\Omega - \lambda^{(i)} + {\rm i}
\epsilon^{(i)}) (\Omega + \lambda^{(i)} +
{\rm i} \epsilon^{(i)})}\,, \nonumber \\
R_{Y_2F}^{(i)} &=& - \sqrt{\frac{\epsilon^{(i)} \theta^{(i)} m}{2
\hbar}} \frac{\epsilon^{(i)} - {\rm i} \Omega}{(\Omega -
\lambda^{(i)} + {\rm i} \epsilon^{(i)}) (\Omega + \lambda^{(i)} +
{\rm i} \epsilon^{(i)})}\,, \nonumber
\end{eqnarray}
where the (free) optical resonant frequency of the large-scale
interferometer at $\Omega = - \lambda^{(1)} - {\rm i}
\epsilon^{(1)}$ is determined by
\begin{eqnarray}
\lambda^{(1)} &=& \gamma_o \frac{2 \rho_{\rm SR} \sin(2 \phi)}{1 +
\rho_{\rm SR}^2 + 2
\rho_{\rm SR} \cos(2\phi)}\,, \nonumber \\
\epsilon^{(1)} &=& \gamma_o \frac{1 - \rho_{\rm SR}^2}{1 + \rho_{\rm
SR}^2 + 2 \rho_{\rm SR} \cos(2\phi)}\,. \nonumber
\end{eqnarray}

As already mentioned, the second carrier does not resonate in the
arm cavities and therefore the local meter is just equivalent to a
interferometer configuration without cavities in the arms. Thus,
in Eq.\,(\ref{eombs}) we only take into account the forces on the
BS due to field fluctuations around the second carrier, in the
same way as in Ref.~\cite{HSD2004}: the first two terms in the
bracket on the right-hand side of Eq.\,(\ref{eombs}) are due to
dark-port fluctuations around the second carrier, while the third
and fourth term, given by
\begin{eqnarray}
\hat F_{\rm BP}^{(1)} & = & \gamma_0 \frac{L^{(1)}
\sqrt{\theta^{(1)}m\hbar(1-\rho_{\rm
PR}^2)\gamma_0}}{\sqrt{2}c(-\gamma_0 (1-\rho_{\rm PR})+{\rm
i}(1+\rho_{\rm PR})\Omega)}\ b_1^{(1)}\,, \nonumber\\
\hat F_{\rm BP}^{(2)} & = & \sqrt{\frac{\theta^{(2)}m L^{(2)}
\hbar (1 + \rho_{\rm PR})}{2 c (1 - \rho_{\rm PR})}} \
b_1^{(2)}\,, \nonumber
\end{eqnarray}
are forces due to bright-port fluctuations, where $b_1^{(i)}$ are
the amplitude quadrature of fluctuations around the first and
second carrier, at the input port. Forces due to fluctuations
around the first carrier are usually negligible, because the
intensity of the first carrier at the beam splitter is lower than
that of the second carrier; in addition, fluctuations associated
with the first carrier also do not build up as much as those
associated with the second carrier, both in common and in
differential mode.

In Eq.\,(\ref{eom1}), we make the approximation that the first
carrier only senses the cavity length, $\hat{x}_{\rm
ETM}-\hat{x}_{\rm ITM}$, ignoring the slight difference between
its sensitivities to ITM and ETM, as well as motion of the BS.  In
Eq.\,(\ref{eom2}), the second carrier only senses the ITM and BS
motions, since it does not enter the arm cavities.

\begin{figure*}
\centerline{\includegraphics[height=8.5cm,angle=-90]{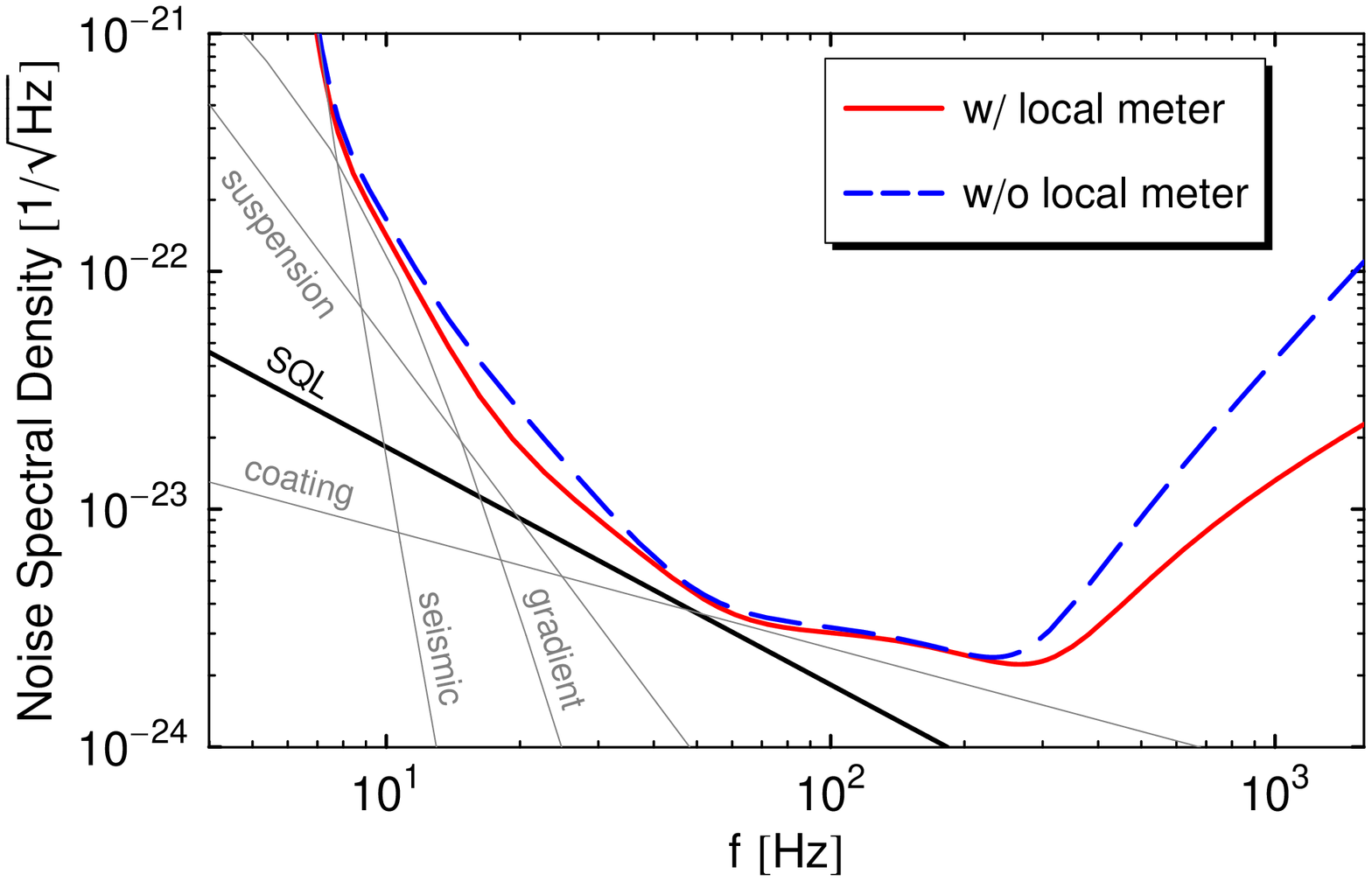}\hspace*{1cm}
            \includegraphics[height=8.5cm,angle=-90]{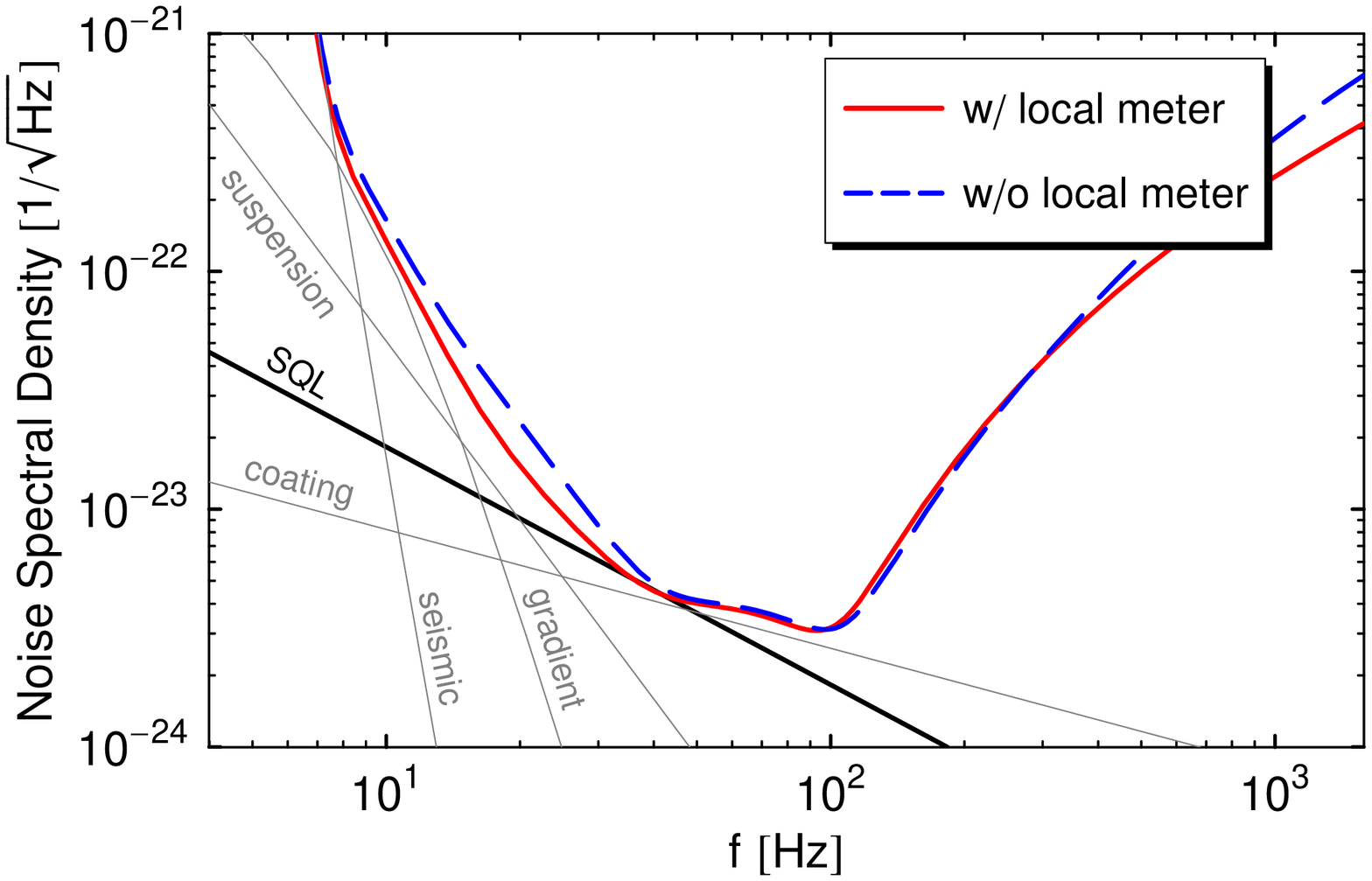}}
\centerline{\includegraphics[height=8.5cm,angle=-90]{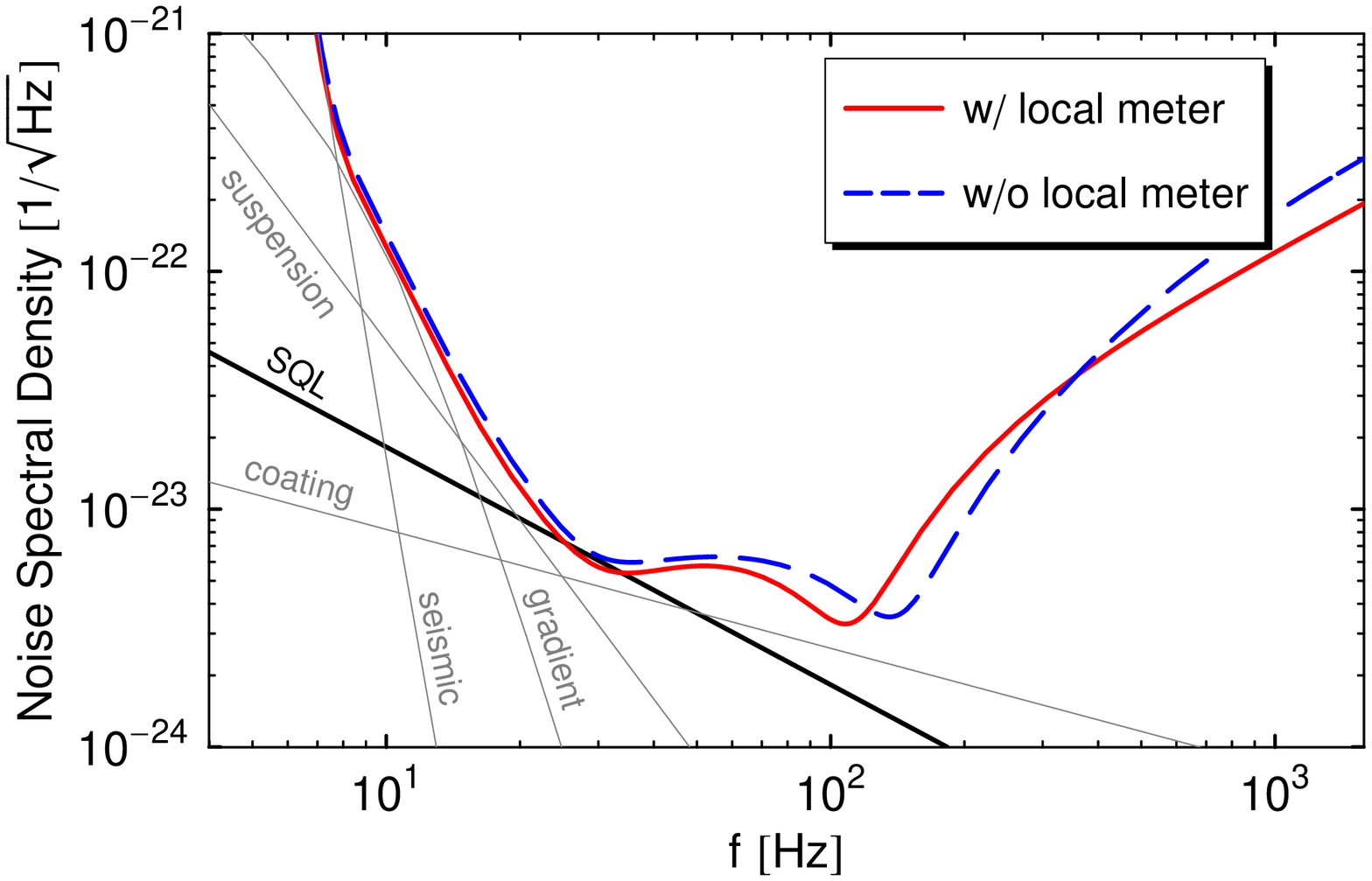}}
\caption{Noise curves for the scheme with local readout (power of
2\textsuperscript{nd} carrier fixed to $P^{(2)}=4 \ {\rm kW}$) and
without local readout both optimized for binary systems with total
mass $M = 2.8 \ M_\odot$ (upper left), $M = 40 \ M_\odot$ (upper
right) and $M = 120 \ M_\odot$ (lower). Special parameters used
for optimizations are given in Tab.\,\ref{improvement} and all
others in Tab.\,\ref{definitions}. Here classical noise (grey
lines) is included. Single contributions of the classical noise
are labeled according to their appearance: suspension thermal
noise results from the fluctuations in the suspension system;
seismic noise is due to motion of the ground; thermal fluctuations
in the coating dominates the one in the substrate; gravity
gradient noise accounts for time-changing Newtonian gravitational
forces.} \label{noisecurve4}
\end{figure*}

The operators $\hat \xi_{\rm ITM}$, $\hat \xi_{\rm ITM}$ and $\hat
\xi_{\rm BS}$ model the classical noise at ITM, ETM and BS,
respectively. We assume that they are uncorrelated but all have
the same spectrum, namely, one fourth of the classical noise
spectrum generally expected for the differential mode of motion.
\begin{table}
\begin{center}
\begin{tabular}{|c|l|l|}
  \hline
  Symbol & physical meaning & value \\
  \hline
  $m$ & single mirror mass & $40 \ {\rm kg}$ \\
  $m_{\rm BS}$ & beam splitter mass & $40 \ {\rm kg}$ \\
  $c/\omega_0^{(1)}$ & laser wavelength of 1\textsuperscript{st} carrier & $1064 \ {\rm nm}$ \\
  $P^{(1)}$ & circulating power of 1\textsuperscript{st} carrier & $0.1 \dots 0.8 \ {\rm MW}$ \\
  $L^{(1)}$ & large-scale interferometer arm length  & $4 \ {\rm km}$ \\
  $\rho_{\rm PR}$ & power-recycling mirror reflectivity & $\sqrt{0.94}$\\
  $\phi$ & detuning phase for 1\textsuperscript{st} carrier & $0 \dots \pi$ \\
  $\rho_{\rm SR}$ & signal-recycling mirror reflectivity & $\sqrt{0.93}$ \\
  $\gamma_o$ & cavity half bandwidth for 1\textsuperscript{st} carrier & $2 \pi \ 15 \ {\rm Hz}$ \\
  $\zeta^{(1)}$ & detection angle for 1\textsuperscript{st} carrier & $0 \dots \pi$ \\
  $c/\omega_0^{(2)}$ & laser wavelength of 2\textsuperscript{nd} carrier & $1064 \ {\rm nm}$ \\
  $P^{(2)}$ & circulating power of 2\textsuperscript{nd} carrier & $0 \dots 16 \ {\rm kW}$ \\
  $L^{(2)}$ & local meter arm length & $15 \ {\rm m}$ \\
  $\lambda^{(2)}$ & detuning for 2\textsuperscript{nd} carrier & $0 \ {\rm Hz}$ \\
  $\epsilon^{(2)}$ & cavity half bandwidth for 2\textsuperscript{nd} carrier & $2 \pi \ 4 \ {\rm kHz}$ \\
  $\zeta^{(2)}$ & detection angle for 2\textsuperscript{nd} carrier & $0$ \\
  \hline
\end{tabular}
\caption{Technical data and parameter values for large-scale
interferometer and local meter used throughout the calculations.
Note that we have defined $\rho_{\rm SR}$ with respect to the
first carrier while the local meter requires a different
reflectivity in order to achieve the bandwidth
$\epsilon^{(2)}$.}\label{definitions}
\end{center}
\end{table}
By using the following only non-vanishing correlation functions
\begin{eqnarray} \label{spec}
\langle \hat a_k^{(i)} (\Omega)\ (\hat a_l^{(j)})^\dagger (\Omega')
\rangle_{{\rm sym}} &=& \pi \ \delta (\Omega - \Omega') \
\delta_{ij} \ \delta_{kl}\,, \nonumber \\
\langle \hat b_k^{(i)} (\Omega)\ (\hat b_l^{(i)})^\dagger
(\Omega') \rangle_{{\rm sym}} &=& \pi \ \delta (\Omega - \Omega')
\ \delta_{kl} \ S_{\rm l}^{(i)} (\Omega)\,, \nonumber \\
\langle \hat \xi_{\rm ITM} (\Omega)\ (\hat \xi_{\rm ITM})^\dagger
(\Omega') \rangle_{{\rm sym}} &=& 2 \pi \
\delta (\Omega - \Omega') \ S_{\rm cl} (\Omega)\,, \nonumber \\
\langle \hat \xi_{\rm ETM} (\Omega)\ (\hat \xi_{\rm ETM})^\dagger
(\Omega') \rangle_{{\rm sym}} &=& 2 \pi \
\delta (\Omega - \Omega') \ S_{\rm cl} (\Omega)\,, \nonumber \\
\langle \hat \xi_{\rm BS} (\Omega)\ (\hat \xi_{\rm BS})^\dagger
(\Omega') \rangle_{{\rm sym}} &=& \pi \ \delta (\Omega - \Omega')
\ S_{\rm cl} (\Omega) \,,
\end{eqnarray}
we obtain the single-sided noise spectral densities. Here
$S^{(i)}_{\rm l} (\Omega)$ is the spectrum of technical input
laser noise while $S_{\rm cl} (\Omega)$ characterizes the spectrum
of all the other classical noise sources. In further calculations
we will assume amplitude laser noise to be white and ten times in
power above shot noise level. For other classical noise sources,
we use the current noise budget of Advanced LIGO, as given in
\emph{Bench} \cite{bench}; contributions such as suspension
thermal noise, seismic noise, thermal fluctuations in the coating
and gravity gradient noise are presented in
Fig.\,\ref{noisecurve4}.

Note that we can obtain two input-output relations from the
equation of motions in Eq.\,(\ref{eomitm})-(\ref{eom2}) and write
them in the following compact form
\begin{equation}
\hat y^{(1)} = \vec n_1^T \ \vec \nu + s_1 \ h \,,\quad
 \hat
y^{(2)} = \vec n_2^T \ \vec \nu + s_2 \ h\,,
\end{equation}
where $\vec \nu^T = (\hat a_1^{(1)}, \hat a_2^{(1)}, \hat
a_1^{(2)}, \hat a_2^{(2)}, \hat b_1^{(1)}, \hat b_1^{(2)}, \hat
\xi_{\rm ITM}, \hat \xi_{\rm ETM}, \hat \xi_{\rm BS})$ and $T$
denotes transposed. Here the two vectors $\vec n_{1,2}$ are the
linear transfer functions from the noise channels $\vec \nu$ into
the two output channels, while the two functions $s_{1,2}$ are the
linear transfer functions from the signal, i.e. the GW strain $h$,
into the output channels.

\subsection{Combined sensitivity}
\label{sec:eqm:sn}

Now we seek for a linear combination of the two output channels,
$\hat y^{(1)}$ and $\hat y^{(2)}$,
\begin{equation}\label{combiout}
\hat y = K_1 (\Omega) \ \hat y^{(1)} + K_2 (\Omega) \ \hat
y^{(2)}\,,
\end{equation}
which has optimal sensitivity to gravitational waves.  In this
optimization, we only consider the signal-referred noise spectral
density of $\hat y$,
\begin{equation} \label{snd}
S_h (\Omega) =
\frac{\left(\begin{array}{cc} K_1 & K_2 \\
\end{array} \right)
\mathbf{N}
\left(\begin{array}{c} K_1^* \\
K_2^* \\ \end{array}\right)}{\left( \begin{array}{cc} K_1 & K_2 \\
\end{array} \right)\mathbf{S} \left(\begin{array}{c} K_1^* \\
K_2^* \\ \end{array}\right)}
\end{equation}
with
\begin{equation}
\mathbf{N} \equiv
\left[\begin{array}{c} \vec n_1^T   \\
\vec n_2^T
\end{array}\right]\left[
\begin{array}{ccccc}
\mathbbmss{1}_4 \\
& S_{\rm l}^{(1)} \\
& & S_{\rm l}^{(2)} \\
& & & 2 S_{\rm cl}\mathbbmss{1}_2 \\
& & & & S_{\rm cl}
\end{array}
\right]
\left[\begin{array}{cc}  \vec n_1^* &  \vec n_2^* \end{array}\right]
\end{equation}
and
\begin{equation}
\mathbf{S} \equiv
\left(\begin{array}{cc}  s_1  s_1^* &
 s_1  s_2^*  \\  s_2 s_1^*  &  s_2  s_2^*
\\ \end{array}\right)\,,
\end{equation}
where $\mathbbmss{1}_k$ stands for a $k$-dimensional identity
matrix. One way of obtaining the minimum noise is to impose the
constraint that the value of the denominator always remains unity,
and minimize the numerator under this constraint. Note that an
overall rescaling of the vector $(K_1,K_2)$ does not affect $S_h$.
The resulting minimum noise is one over the bigger eigenvalue of
the 2-by-2 matrix
\begin{equation}
\mathbf{M} \equiv \mathbf{N}^{-1} \mathbf{S}\,,
\end{equation}
with the corresponding eigenvector providing the optimal filters
$(K_1,K_2)$.
\begin{figure*}
\begin{tabular}{ccc}
\includegraphics[height=8.6cm,angle=-90]{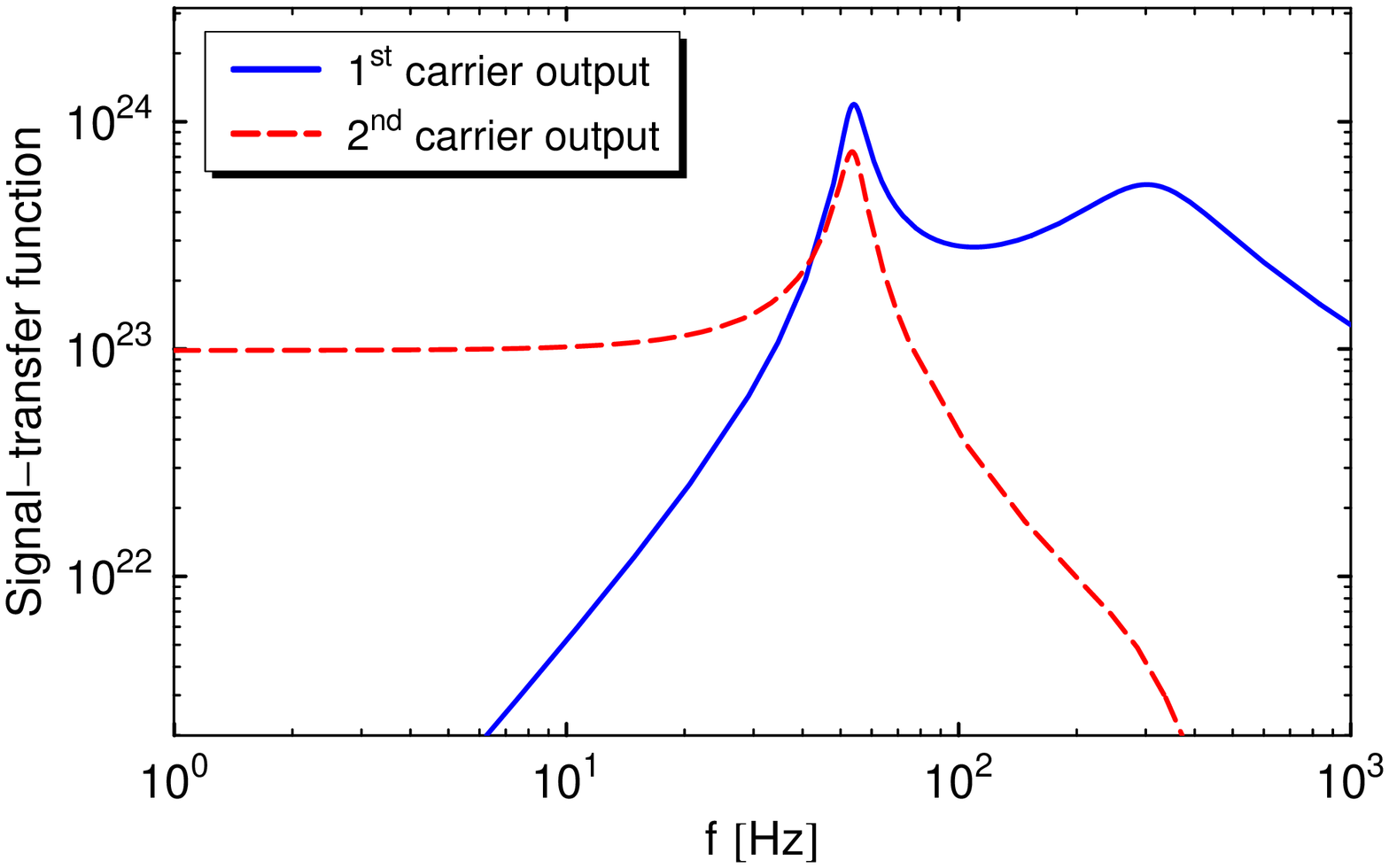}
&
\hspace{0.5cm}
&
\includegraphics[height=8.6cm,angle=-90]{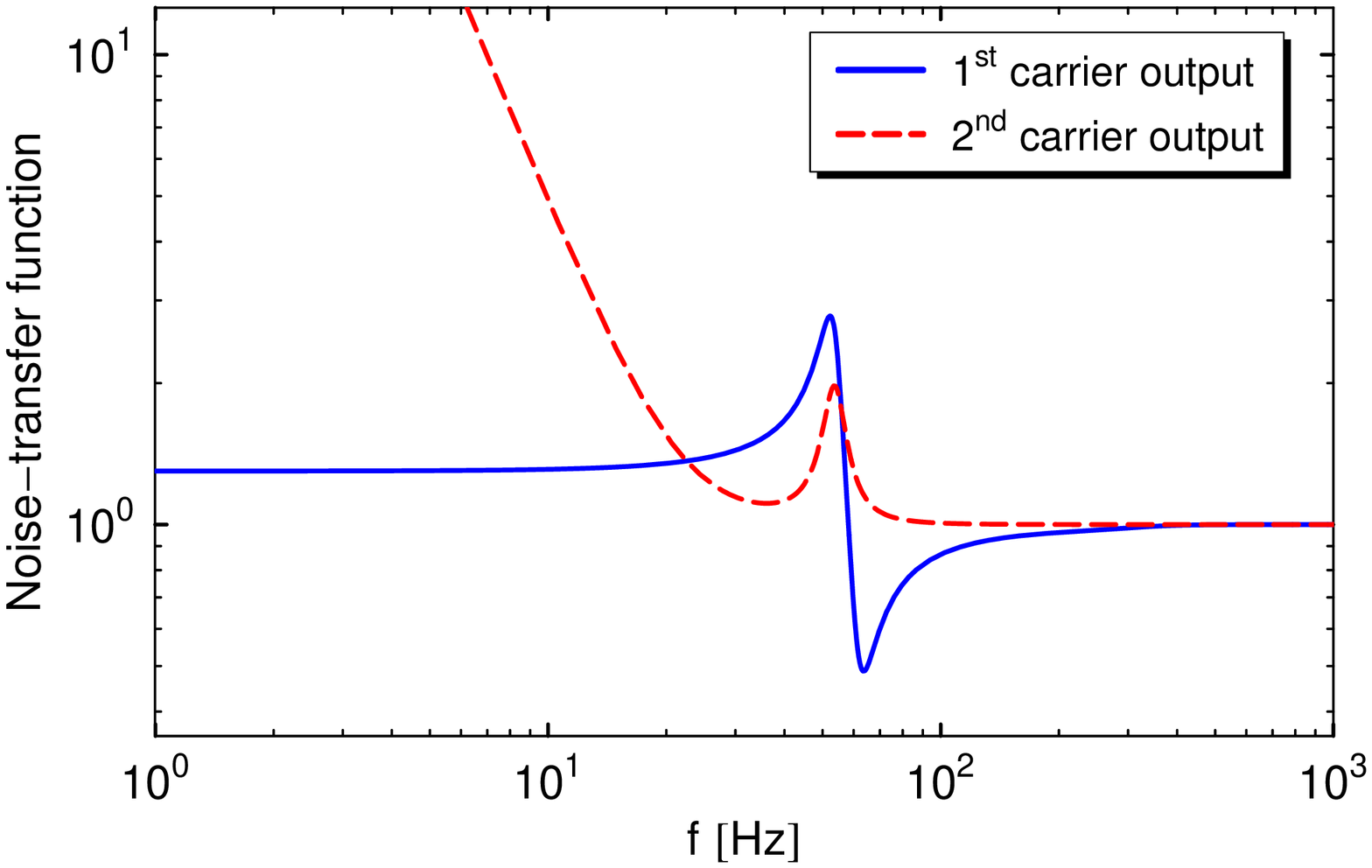}
\end{tabular}
\caption{Example for the signal (left panel) and noise (right
panel) transfer functions in a signal-recycled Michelson
interferometer with two carriers and double-readout, for a
configuration with the parameters as given in
Tab.\,\ref{definitions} but $\zeta^{(1)} = 0$, $\phi=\pi/2-0.014
\pi$ and $P^{(1)}=800\,{\rm kW}$).} \label{strans}
\end{figure*}
\begin{figure*}
\begin{tabular}{ccc}
\includegraphics[height=8.6cm,angle=-90]{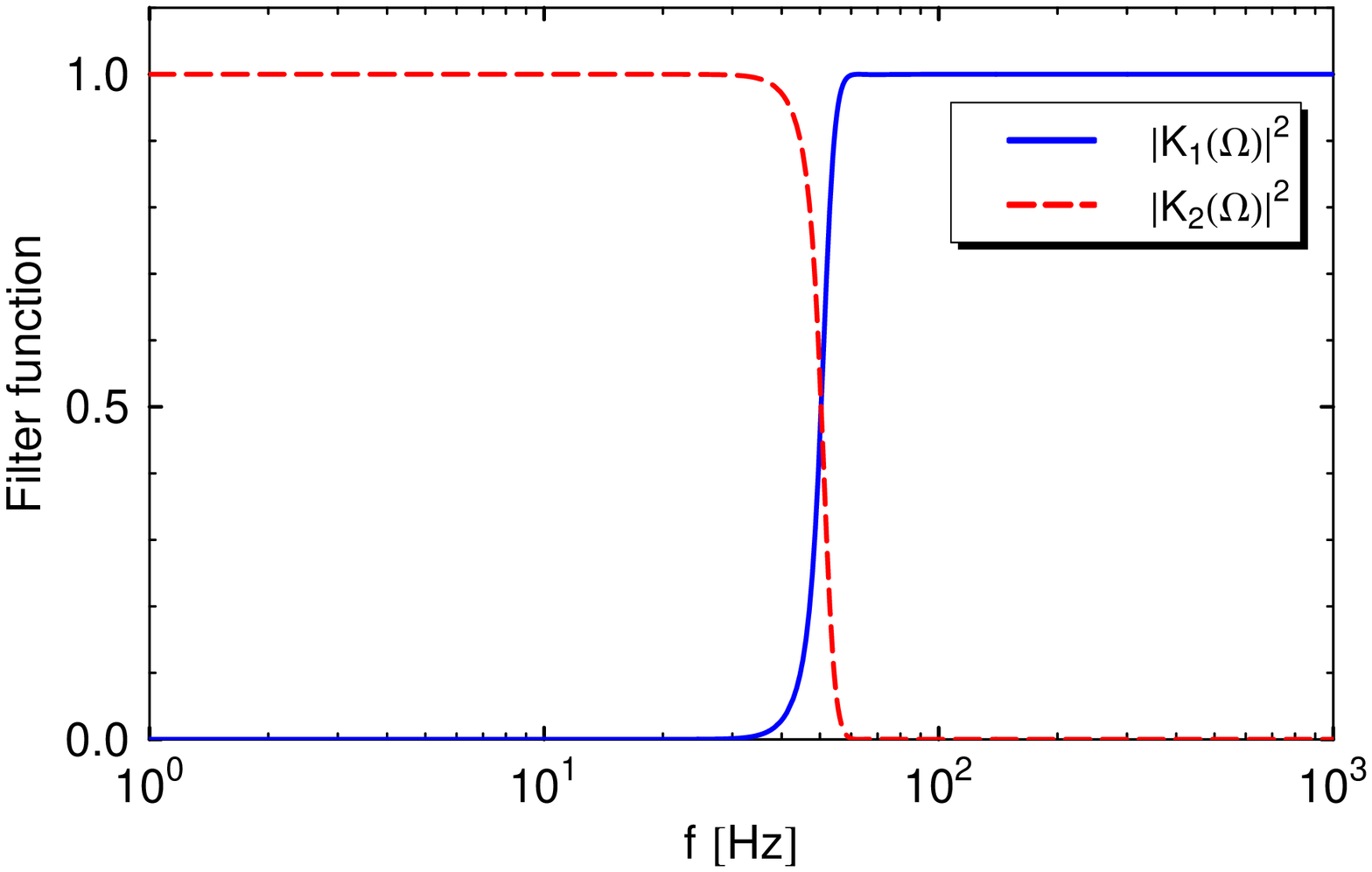}
&
\hspace{0.5cm}
&
\includegraphics[height=8.6cm,angle=-90]{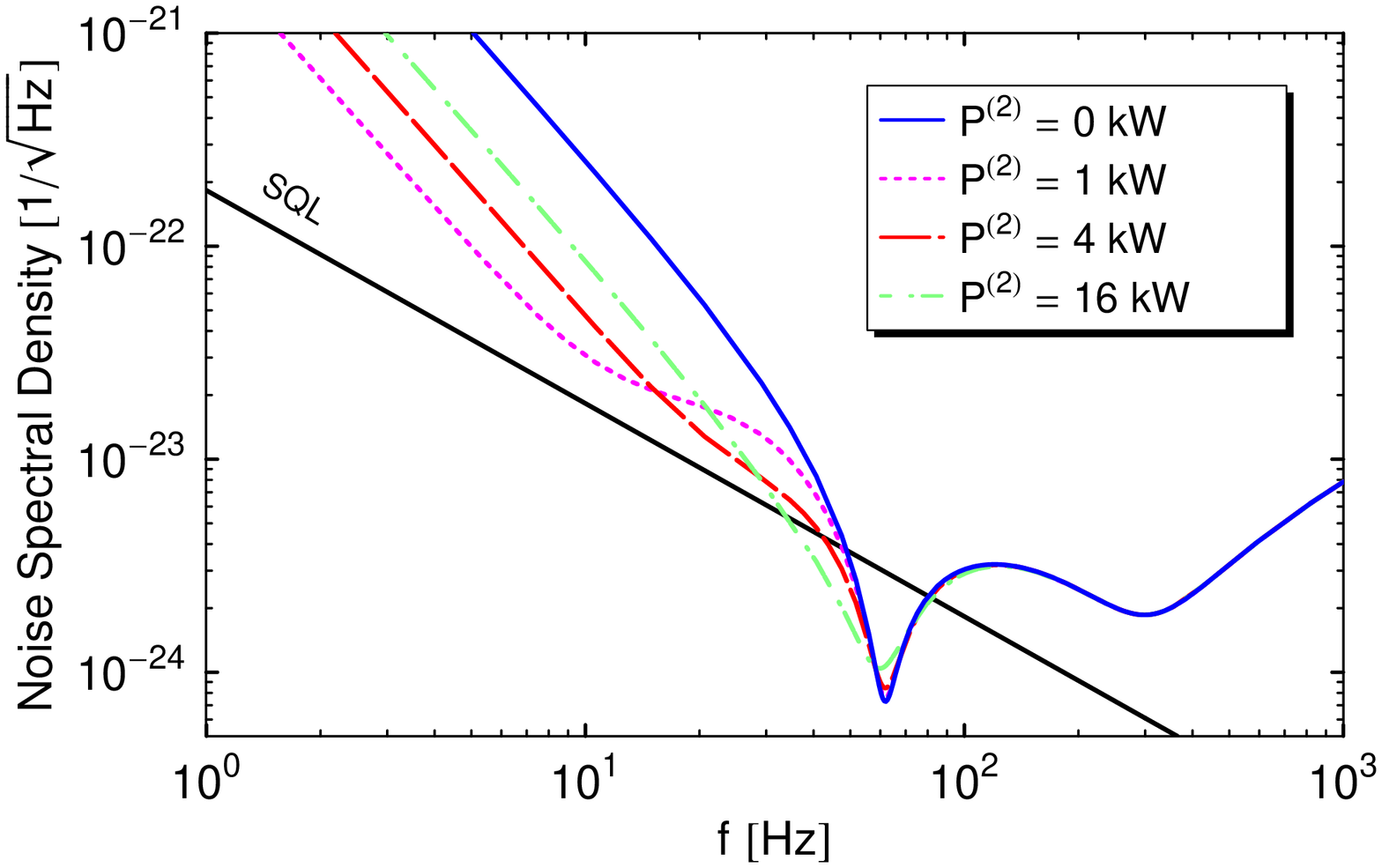}
\end{tabular}
\caption{Left panel: filter functions $K_1$ and $K_2$, for the
same configuration as in Fig.\,\ref{strans}. Here each filter
function is rescaled such that it gives the percentage of how much
GW strain it feeds into the combined output. Right panel: quantum
noise curves for our proposed scheme with different powers of
2\textsuperscript{nd} carrier. Again phase quadrature readout
$\zeta^{(1)} = 0$, signal-recycling cavity detuning phase
$\phi=\pi/2-0.014 \pi$ and power $P^{(1)}=800\,{\rm kW}$ of
1\textsuperscript{st} carrier are used.} \label{noisecurve1}
\end{figure*}

We now illustrate the local readout scheme using the following
configuration: the parameters are given in Tab.\,\ref{definitions}
as well as phase quadrature readout $\zeta^{(1)} = 0$,
signal-recycling cavity detuning phase $\phi=\pi/2-0.014 \pi$ and
power $P^{(1)}=800\,{\rm kW}$ of the first carrier are used. In
Fig.\,\ref{strans} we plot the individual signal- and
noise-transfer functions of the first and second carriers, for the
configuration with $P^{(2)}=4\,{\rm kW}$. As we can see from these
plots, the first carrier mainly senses frequencies above the
optical-spring resonance with signal-transfer function suppressed
at lower frequencies by the optical spring; the second carrier
offers complementary sensitivity for frequencies below the
optical-spring resonance, when the ITM is dragged together with
the ETM by the optical spring. As a consequence, as we see in the
left panel of Fig.\,\ref{noisecurve1}, at frequencies above the
optical-spring resonance, the optimal combination depends mostly
on the first readout, while at frequencies below the
optical-spring resonance, the optimal combination depends mostly
on the second readout.

Noise curves with optimal filters are plotted for different powers
of the second carrier (0\,kW, 1\,kW, 4\,kW and 16\,kW) in the
right panel of Fig.\,\ref{noisecurve1} where only quantum noise is
taken into account. This plot illustrates that the local readout
scheme can directly improve the sensitivity only below the
optomechanical resonance frequency. It turns out that $4 \ {\rm
kW}$ in each arm of the local meter already gives a remarkable
increase in sensitivity. In the following studies we fix
$P^{(2)}=4 \ {\rm kW}$.

One could imagine that the combination of a signal-recycled
Michelson interferometer with a local readout may indirectly help
improving the sensitivity at high frequencies or increasing the
detection bandwidth, once an overall optimization to a broadband
source is performed. The underlying effect is that the sensitivity
of the large-scale interferometer can be shifted to higher
frequencies by choosing its detection angle to be closer to the
phase quadrature while the local meter helps to maintain
sensitivity at low frequencies. This will be studied more
carefully in Sec.~\ref{sec:SNR}.

\subsection{Control}
\label{sec:eqm:ctrl}

As it has been shown in Refs.~\cite{BUCH2001,BUCH2002,BUCH2003}
the optical spring introduces an instability, which must be
stabilized using a feedback control system.  In single-readout
systems, it is easy to show that such a control system does not
give rise to any fundamental change in our GW sensitivity
\cite{BUCH2001,BUCH2002,BUCH2003}, intuitively because signal and
noise are fed back with the same proportion onto the test masses.
Our double-readout system is more complex, but the same intuition
still applies. If we denote $\vec x \equiv (\hat x_{\rm ITM},\hat
x_{\rm ETM},\hat x_{\rm BS})^T$ and  $\vec y \equiv (\hat
y^{(1)},\hat y^{(2)})^T$, the Eqs.~(\ref{eomitm})--(\ref{eom2})
can be written schematically as:
\begin{eqnarray}
\label{eomgen1}
\vec x &=& \mathbf{A}(\Omega) \vec x + \mathbf{B}(\Omega)\vec \nu
+ \vec C(\Omega) h + \mathbf{D}(\Omega)\vec y\,, \\
\label{eomgen2}
\vec y &=& \mathbf{F}(\Omega) \vec x + \mathbf{G}(\Omega) \vec\nu\,.
\end{eqnarray}
Here matrix  $\mathbf{A}$ describes mirror dyanmics, matrix
$\mathbf{B}$ describes how the noise sources in $\vec \nu$ are
applied as forces onto the mirrors, vector $\vec C$ describes how
GW signal $h$ directly influences the mirrors, $\mathbf{F}$
describes how the output channels $\vec y$ sense the various
motions $\vec x$, $\mathbf{G}$ describes sensing noise in $\vec
y$, and finally $\mathbf{D}$ describes the feedback.  Solving
Eqs.~(\ref{eomgen1}) and (\ref{eomgen2}) jointly, we obtain
\begin{equation}
\label{eomygen} \vec y =[\mathbbmss{1}_2 -
\mathbf{H}\mathbf{D}]^{-1}
\left[[\mathbf{H}\mathbf{B}+\mathbf{G}]\vec\nu +\mathbf{H}\vec C
h\right]\,,
\end{equation}
where we have defined $\mathbf{H} \equiv
\mathbf{F}(\mathbbmss{1}_2-\mathbf{A})^{-1}$.  In
Eq.\,(\ref{eomygen}) the only dependence of $\vec{y}$ on the
control system is through $\mathbf{D}$, which only appears in the
first factor on the right-hand side. The optimal sensitivity,
which is obtained by maximizing signal-referred noise spectrum of
$(K_1\;K_2) \ \vec y$, is then clearly invariant with respect to
changes in $\mathbf{D}$.

\section{Improvements in Advanced LIGO sensitivity}
\label{sec:SNR}

\subsection{Matched-filtering signal-to-noise ratio}

To quantify the astrophysical merit of various configurations, we
will calculate the improvement in the matched-filtering
signal-to-noise ratio (SNR) or the detectable distance for a given
threshold SNR, respectively, for inspiral waves from compact
binary systems. For a known waveform (in the frequency domain)
$h(f)$, the optimal SNR achievable by correlating the data with a
known template is
\begin{equation}\label{SNR}
\rho = 2 \sqrt{\int_0^{\infty}{\rm d}f \ \frac{| h(f)|^2}{S_h(f)}}
\end{equation}
where $S_h(f)$ is the single-sided noise spectral density. For
compact binary objects, the lowest Post-Newtonian approximation
gives  (see, e.g., \cite{DIS2000})
\begin{equation}
\label{hfinspiral}
|h(f)|=\frac{G^{5/6} \mu^{1/2} M^{1/3}}{\sqrt{30} \pi^{2/3} c^{3/2}
D} \ f^{-7/6} \ \Theta(f_{\max}-f)
\end{equation}
with
\begin{equation}
M=(M_1+M_2)\quad {\rm and}\quad\mu=\frac{M_1 M_2}{M_1+M_2}
\end{equation}
where $\mu$, $M$, $M_1$ and $M_2$ are the reduced, total and
single masses of the binary and $D$ is the distance from the
source to the detector. Here the amplitude is the one where rms
average over all directions is already taken into account. There
is an upper cutoff frequency, $f_{\rm max}$, in
Eq.\,(\ref{hfinspiral}) beyond which the systems undergoes a
transition from adiabatic inspiral into non-adiabatic merger, and
Eq.\,(\ref{hfinspiral}) is no longer a valid approximation.  This
frequency is usually taken to be the GW frequency at the last
stable circular orbit given, for a test mass in a Schwarzschild
space time with mass $M$
\begin{equation}
f_{\rm max} \approx 4400 \ {\rm Hz}\left(\frac{M_\odot}{M}\right)\,.
\end{equation}
A lower cut-off frequency $f_{\rm min}$  should also be applied to
the integration in  Eq.\,(\ref{SNR}), below which it is no longer
possible to treat the system as stationary. We take $f_{\rm min}
\approx 7 \ {\rm Hz}$. Considering binaries of averaged
orientation the observable distance for a given SNR $\rho_0$
reaches
\begin{equation}
D = \sqrt{\frac{2}{15}} \frac{G^{5/6}\mu^{1/2} M^{1/3}}{\pi^{2/3}
c^{3/2} \rho_0} \sqrt{\int_{f_{\rm min}}^{f_{\rm max}}{\rm d}f \
\frac{f^{-7/3}}{S_h(f)}}\,.
\end{equation}
In this paper, we assume event rate to be proportional to the cube
of detectable distance, i.e.,
\begin{equation}
\mathcal{R} \propto D^3\,.
\end{equation}

\subsection{Improvement in the event rate}

\begin{table*}
\begin{center}
\begin{tabular}{|c|c|c|c|c|c|c|c|}
  \hline
  $M/M_\odot$ & \multicolumn{3}{|c|}{ \ optimization parameters w/ local meter \ }
   & \multicolumn{3}{|c|}{ \ optimization parameters w/o local meter \ } & improvement \\
  \cline{2-4} \cline{5-7}
  & $P^{(1)}$ in kW & $\phi$ in radian & $\zeta^{(1)}$ in radian & $P^{(1)}$ in kW
  & $\phi$ in radian & $\zeta^{(1)}$ in radian & in event rate \\
  \hline
  2.8 & 800 & 0.48 \ $\pi$ & 0.7 \ $\pi$ & 800 & 0.48 $\pi$ & 0.49 $\pi$ & 29 \% \\
  20 & 450 & 0.47 \ $\pi$ & 0.58 \ $\pi$ & 500 & 0.48 $\pi$ & 0.48 $\pi$ & 28 \% \\
  30 & 250 & 0.46 \ $\pi$ & 0.46 \ $\pi$ & 200 & 0.46 $\pi$ & 0.49 $\pi$ & 30 \% \\
  40 & 150 & 0.45 \ $\pi$ & 0.43 \ $\pi$ & 150 & 0.45 $\pi$ & 0.46 $\pi$ & 33 \% \\
  80 & 100 & 0.45 \ $\pi$ & 0.38 \ $\pi$ & 100 & 0.45 $\pi$ & 0.46 $\pi$ & 44 \% \\
  120 & 100 & 0.46 \ $\pi$ & 0.32 \ $\pi$ & 100 & 0.47 $\pi$ & 0.41 $\pi$ & 42 \% \\
  160 & 110 & 0.47 \ $\pi$ & 0.25 \ $\pi$ & 100 & 0.47 $\pi$ & 0.30 $\pi$ & 45 \% \\
  200 & 110 & 0.48 \ $\pi$ & 0.25 \ $\pi$ & 100 & 0.48 $\pi$& 0.27 $\pi$ & 48 \% \\
  \hline
\end{tabular}
\caption{Parameters used when optimizing our proposed
double-readout scheme and the usual Advanced LIGO like
configuration each for different binary systems. Last column gives
the improvement in the event rate for our proposed scheme compared
to the usual scheme both optimized for the given equally
distributed total binary mass. Reasonable errors in $\phi$ and
$\zeta^{(1)}$ may decrease the event rate -- but not more than
1\%. }\label{improvement}
\end{center}
\end{table*}
The tools reviewed in the previous subsection enable us to
optimize a specific interferometer configuration for given binary
inspirals by maximizing its SNR with respect to certain
interferometer parameters. Note that we now take also classical
noise into account as it is indicated by the grey lines in
Fig.\,\ref{noisecurve4}. In this paper we assume that Advanced
LIGO refers to a signal-recycled interferometer without local
readout and optimized for neutron star - neutron star (NS-NS)
binary systems, i.e. binary system with $M=(1.4+1.4) \ M_{\odot}$.
We then vary the optical power $P^{(1)}$, detuning $\phi^{(1)}$
and detection angle $\zeta^{(1)}$ in such a way that the SNR of
the signal-recycled interferometer without local readout is
maximized for the total mass of a given binary system in
Tab.\,\ref{improvement}. When we optimize our scheme,  we maximize
the SNR varying the same set of parameters of the large-scale
interferometer (cf. Tab.\,\ref{improvement}) but with imposing a
fixed power for the second carrier ($P^{(2)}=4\,{\rm kW}$),
requiring the second carrier to be resonant in the
signal-recycling cavity ($\lambda^{(2)}=0$), and fixing a
detection quadrature phase  of $\zeta^{(2)}=0$ (i.e., detecting
the phase quadrature). Such a prescription is justified, because a
local meter with such a short arm length, low power and finesse
(as we have chosen) is mostly dominated simply by  shot noise.

If we compare the two schemes with and without an added local
meter at the binary mass they are optimized for we find moderate
improvement in event rates (cf. last column in
Tab.\,\ref{improvement}). The improvement increases for higher
binary masses since our scheme helps to enhance sensitivity mainly
at low frequencies. Such a moderate improvement has been limited
mainly due to low-frequency classical noise.

\begin{figure}
\includegraphics[height=8.5cm,angle=-90]{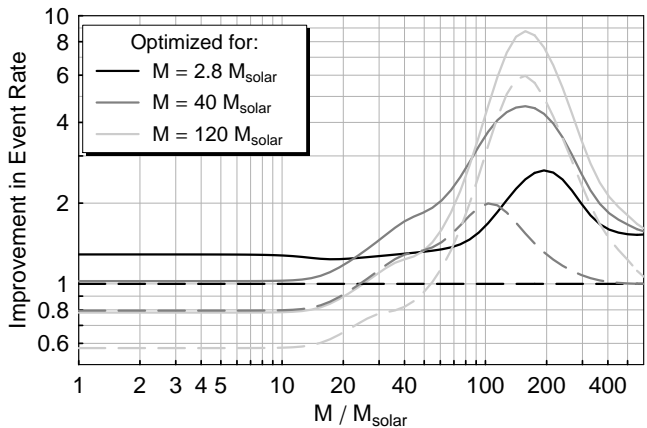}
\caption{Improvement in the event rate compared to Advanced LIGO
versus total binary mass with fixed optimization parameters for
each curve. Signal-recycled interferometer with (solid lines) and
without (dashed lines) local readout are optimized for three
different binary masses. Power of 2\textsuperscript{nd} carrier is
fixed to $P^{(2)}=4 \ {\rm kW}$.} \label{SNR1}
\end{figure}

The advantage of the local readout scheme can be appreciated
better when we realize that there are different populations of
likely sources (e.g., binary total mass $M$ can reside in a range,
$\mathcal{M}$), whose signals extend to different frequency bands.
We need to investigate how good a configuration optimized for a
particular system with total mass $M$ would perform for other
possible masses in $\mathcal{M}$.  In this paper, we consider
$\mathcal{M} = \left[M_{\odot}, \ 630\,M_{\odot}\right]$ with
maximum mass determined by the condition $f_{\rm max} =f_{\rm
min}$. In Fig.\,\ref{SNR1}, we show the improvements in event
rates (with respect to Advanced LIGO baseline, optimized for NS-NS
binaries) obtainable by Advanced LIGO configurations (solid lines)
and double-readout configurations (dashed lines) for binaries with
$M\in\mathcal{M}$, when the configurations are optimized
specifically for $M=2.8\,M_\odot$ (black), $40\,M_\odot$ (dark
gray) and $120\,M_\odot$ (light gray).  In
Fig.\,\ref{noisecurve4}, we show the corresponding noise spectral
densities of these configurations, together with classical noise.
Figures~\ref{SNR1} and \ref{noisecurve4} provides us with at least
two possible applications of the double-readout scheme.

{\it Detector with broader frequency band.}  The sensitivity of
the double-readout configuration optimized for $2.8\,M_\odot$
systems (solid curve on the upper left panel of
Fig.\,\ref{noisecurve4}) is broader in band and globally better
than the baseline design of Advanced LIGO (dashed curve in the
same figure), particularly at higher frequencies; this
demonstrates that when an overall optimization is performed, the
local readout can indirectly improve sensitivity at higher
frequencies. Although Fig.\,\ref{SNR1} (solid curve) does not show
a significant increase in binary event rates, this configuration
is potentially interesting for detecting other sources above
300\,Hz, for example pulsars and Low-Mass X-ray Binaries.

{\it Detector for intermediate-mass black-hole binaries.} The
double-readout configuration optimized for $40\,M_\odot$ systems
(dark gray curve in Fig.\,\ref{SNR1}) has the same sensitivity to
low-mass binary systems as Advanced LIGO baseline (up to
$M=10\,M_\odot$), while improving event rates for $60\,M_\odot$ --
$300\,M_\odot$ by factors of $2$ -- $4.5$.  This allows us to
build a detector sensitive to the more speculative (yet in some
sense astrophysically more interesting) intermediate-mass
black-hole binaries, without sacrificing sensitivity at low-mass
systems which are more certain to exist.  As we see from dashed
curves in Fig.\,\ref{SNR1}, such broad improvement simultaneously
for systems with different total masses is not achievable by
single-readout Advanced LIGO like configurations.  It is also
interesting to note that this configuration only requires a
circulating power of 150\,kW in the arms.

The improvement in event rate increases significantly for higher
binary masses (cf. gray curve in Fig.\,\ref{SNR1} optimized for
$M=120\,M_\odot$) since the local meter helps to enhance
sensitivity mainly at low frequencies. But if we optimize for such
high masses the sensitivity for lower masses cannot keep up with
Advanced LIGO.

It turns out that our scheme even improves sensitivity in the
low-frequency regime when sensitivity is dominated by classical
noise, as can be seen in Fig.\,\ref{SNR1}, since for high binary
masses the dashed curves meet at a factor of $(4/3)^{3/2}$ below
the solid curves. We explain this factor in the Appendix.

\section{Implementation Issues}
\label{sec:implementation}

In this section we discuss the possibility of implementing this
technique explicitly in the Advanced LIGO detector. In fact, the
so-called central Michelson degree of freedom in the detector,
already to be measured to keep the signal-extraction port of the
interferometer in dark fringe, is exactly what our local readout
scheme proposes to measure. However, sensitivity of the current
Michelson control signal must be improved dramatically in order to
be turned into our regime. We note that more precise measurement
of this Michelson degree of freedom also helps to decrease
control-loop noise, which is shot noise imposed on the control
signal coupling to the main signal due to unavoidable
imbalances~\cite{SMFA2006}.

{\it Optical Power.} In the baseline design, a pair of radio
frequency (RF) sidebands created around the main carrier frequency
will be injected to probe the motion as it is already done in
current detectors. However, the power level of current RF
sidebands is not high enough for our local readout. In the
baseline design, the input power is $125 \ {\rm W}$, which is
amplified to $\sim1.0\ {\rm kW}$ at each ITM, due to power
recycling. Only about 1\% of the power at the input port is pumped
into the RF sidebands that resonate in the power-recycling cavity
but not in the arms. Taking into account the fact that the RF
sidebands do not enter the arm cavities and thus suffer from less
optical losses, the power of the Michelson-control sidebands at
the ITM is currently planned to be $\sim 34\,{\rm W}$. Thus, one
needs to raise the current power by $\sim 120$ times in order to
achieve $P^{(2)}=4\,{\rm kW}$. Another more realistic way of
realization is to use a phase-locked secondary laser with its
frequency shifted by an odd number of half free-spectral ranges
from the primary laser to satisfy the off-resonant condition in
the arms. Furthermore, this sub-carrier should almost be in dark
fringe at the signal-extraction port and should be resonant in
both recycling cavities. To achieve a circulating power of
$P^{(2)}=4\,{\rm kW}$ for the sub-carrier we even need a little
more input power than for the primary laser. But we can hope to
use the higher-power laser for the sub-carrier while the
parametric instability~\cite{BSV2001,BSV2002} in the arm cavity
may limit the power of the primary laser. Indeed, a circulating
power of $P^{(2)}=4\,{\rm kW}$ is only a few times more than the
carrier power of the current GEO detector which has a similar
topology compared to the local meter.

{\it Detection.} Each signal at the dark port should be extracted
with some reference field, which will be another set of RF
sidebands in the RF readout scheme, or DC offset light in the DC
readout scheme. The former one leaks through the dark port via
macroscopic asymmetry in the central Michelson interferometer, and
the latter one leaks through the dark port via microscopic
asymmetry between two arm cavities. Either way, the reference
fields for the carrier and the sub-carrier should be isolated
before the photo-detection, otherwise the reference field which is
not used for the signal extraction will just impose extra shot
noise. One way to solve the problem is to make use of orthogonal
polarizations. Before the photo-detection, the carrier and the
sub-carrier accompanied with the reference fields can be separated
by a polarized beam splitter, which is all-reflective to one
polarization and transmissive to the other. In addition, it is
easy to combine the two beams before injection into the
interferometer without losing the power. An alternative way to the
orthogonal polarizations is to use a cavity that can separate the
beams at different frequencies, where one resonates in the cavity
while the other does not. The cavity, a so-called
output-mode-cleaner, is already planned to be used at the
detection port in Advanced LIGO. In the same way an input mode
cleaner cavity can be used to combine two beams before the
injection into the interferometer.

{\it Alternative configuration.}  One may also place the local
meters around the ETMs. In this case, a single laser beam, which
can be different in frequency from the carrier light, should be
split and brought to each end of the arms so that laser noise can
be cancelled out after taking a subtraction of the two ETMs'
motion measurements. A cavity can be implemented as well as it is
proposed for a radiation-pressure-noise reduction method in
\cite{KMS2001,CoHePi2003}. In this way the secondary laser for the
local readout does not need such high power and there is no
concern of a heat problem at the BS and the ITMs. However, in this
case much more additional optical components are required to
realize this configuration.

\section{Conclusion}
\label{sec:conclusion}

Motivated by the optical-bar schemes \cite{BGK1997} and
quantum-locking schemes \cite{KMS2001,CoHePi2003}, we have
proposed injecting a second laser beam into detuned
signal-recycled Michelson interferometers, sensing the
differential motion of the input mirrors, and improving
low-frequency sensitivities of these interferometers, currently at
low frequencies being limited by the rigidity of the optical
spring. We derived the optimal combined sensitivity of this
double-readout scheme, and demonstrated that this optimal
sensitivity is invariant with respect to the application of a
feedback control scheme.

Taking into account the current classical noise budget of Advanced
LIGO, as well constraints on optical power, we performed an
optimization of our double-readout schemes toward the detection of
compact binary inspirals. This scheme is shown either to be able
to broaden the detection band and (indirectly) significantly
improving high-frequency sensitivities, or to allow the detection
of intermediate-mass black-hole binaries with a broad frequency
range without sacrificing sensitivity to neutron-star binaries and
stellar-mass black-hole binaries.

We also discussed briefly how the sensing of the Michelson degree
of freedom in the currently plan of Advanced LIGO can be made
dramatically more sensitive and turned into our local readout
scheme.

Finally, we would like to point out that this scheme should be
further investigated as a candidate design for third-generation
detectors, possible in conjunction with the injection of squeezed
vacuum states~\cite{VCHFDS2006,McKenzie04} into the
interferometer's dark port~\cite{HCCFVDS2003,BUCH2004}.

\begin{acknowledgments}
We thank Stan Whitcomb for very useful discussions. Research of
H.M.-E.\,, K.S.\, and Y.C.\, is supported by the Alexander von
Humboldt Foundation's Sofja Kovalevskaja Programme. Research of
H.R.\, and R.S.\, is supported by the Deutsche
Forschungsgemeinschaft through the EGC programme and the SFB No.
407, respectively. Research of C.L. is supported in part by NSF
grants PHY-0099568 and PHY-0601459.
\end{acknowledgments}

\begin{appendix}

\section{Double-readout scheme dominated by classical noise}

Suppose at low frequencies, sensing noise is negligible, and noise
is dominated by the classical force noise acting on the mirrors.
Then, the first carrier offers the following
\begin{equation}
\hat y^{(1)} \propto \hat \xi_{\rm ETM} - \hat \xi_{\rm ITM} + L
h\,,
\end{equation}
where $\xi_{\rm ETM}$ and $\xi_{\rm ITM}$ are classical noise on
the ITM and the ETM, respectively. The output of the second
carrier is proportional to
\begin{equation}
\hat y^{(2)} \propto \hat \xi_{\rm ETM} + \hat \xi_{\rm ITM} + 2
\sqrt{2} \hat \xi_{\rm BS} + L h\,,
\end{equation}
where $\xi_{\rm BS}$ is the classical noise acting on the BS.
Suppose again that $\xi_{\rm ITM}$, $\xi_{\rm ETM}$, and $\xi_{\rm
BS}$ have independent noise at the same level for ITM and ETM but
half as high for the BS (cf. Eqs.\,(\ref{spec})). We obtain that
the optimal filter uses 3/4 of the output of the large-scale
interferometer and one fourth of the small interferometer in the
units as above. This is in contrast to the optimal filter
functions when only quantum noise is taken into account as in the
left panel of Fig.\,\ref{noisecurve1}. Then the combined output is
given by
\begin{equation}
\hat y \propto \xi_{\rm ETM} - \frac{1}{2} \xi_{\rm ITM} +
\frac{1}{\sqrt{2}} \xi_{\rm BS} + L h\,.
\end{equation}
Then the large-scale interferometer's noise spectral density
versus the optimal noise spectral density reads $2 / \frac{3}{2}$
which gives the factor in Fig.\,\ref{SNR1}. In this way the
double-readout is able to cancel some fraction of the classical
noise.

\end{appendix}

\end{document}